\documentclass[onecolumn]{revtex4-1}
\usepackage{graphicx}
\usepackage{dcolumn}
\usepackage{bm}
\usepackage{amsmath}
\usepackage{amsfonts}
\usepackage{psfrag}
\usepackage{MnSymbol}%
\usepackage{wasysym}%
\usepackage[pdftex,bookmarks,colorlinks=true,linkcolor = blue,urlcolor  = blue,citecolor = black]{hyperref}
 \usepackage[usenames,dvipsnames]{color}
 \usepackage{soul}

\def\bra#1{\mathinner{\langle{#1}|}} 
\def\ket#1{\mathinner{|{#1}\rangle}}

\newcommand{\Eq}[1]{Eq.~(\ref{#1})}

\makeatletter

\begin{document}
\title{Superradiance with local phase-breaking effects}
\author{Nathan Shammah$^{1,2}$, Neill Lambert$^1$, Franco Nori$^{1,3}$, and Simone De Liberato$^2$} 
\affiliation{$^1$ CEMS, RIKEN, Wako-shi, Saitama 351-0198, Japan\\
$^2$ School of Physics and Astronomy, University of Southampton, Southampton, SO17 1BJ, United Kingdom\\
$^3$ Department of Physics, University of Michigan, Ann Arbor, Michigan 48109-1040, USA}

\begin{abstract} 
We study the superradiant evolution of a set of $N$ two-level systems spontaneously radiating under the effect of phase-breaking mechanisms. We investigate the dynamics generated by nonradiative losses and pure dephasing, and their interplay with spontaneous emission. Our results show that in the parameter region relevant to many solid-state cavity quantum electrodynamics experiments, even with a dephasing rate much faster than the radiative lifetime of a single two-level system, a sub-optimal collective superfluorescent burst is still observable.
We also apply our theory to the dilute excitation regime, often used to describe optical excitations in solid-state systems. 
In this regime, excitations can be described in terms of bright and dark bosonic quasiparticles. 
We show how the effect of dephasing and losses in this regime translates into intermode scattering rates and quasiparticle lifetimes.
\end{abstract}
\maketitle
\section{Introduction}
The time a two-level system initially prepared in the excited state $\ket{\uparrow}$ takes to relax into its ground state $\ket{\downarrow}$  emitting a photon is set by the spontaneous emission rate $\gamma_\text{S}$, which quantifies its coupling with the electromagnetic environment. When $N$ identical copies of such a system are prepared in their excited state, $\ket{\uparrow...\uparrow}$, each two-level system emits independently, leading to a total emission rate $N\gamma_\text{S}$. 
Under the condition that the linear size of the ensemble is smaller than the resonance wavelength, subsequent photon emissions then cause a build-up of quantum correlations between different systems, leading after a delay time $t_\text{d}= \frac{\log{N}}{ N\gamma_\text{S}}$ \cite{Dicke54,Gross82} to the emission of light in a short and intense burst of peak intensity $\propto N^{2}$, with a narrow pulse duration $\tau_\text{p}\sim \frac{1}{ N\gamma_\text{S}}$. Such phenomenon, referred to as superfluorescence {or Dicke's superradiance}, is an example of spontaneous symmetry breaking triggered by quantum fluctuations and leading to the emergence of macroscopic quantum correlations in time \cite{Vrehen80}.

Superfluorescence is due to the fully-symmetric coupling with the electromagnetic field. If the system is initially prepared in the totally-excited, maximally-symmetric state $\ket{\uparrow...\uparrow}$, photon emission will only cause transitions within the $N+1$ symmetric states, eventually reaching the ground state $\ket{\downarrow...\downarrow}$. The evolution of the system will thus be bound to the $(N+1)$-dimensional, symmetric subspace of the full $2^N$-dimensional Hilbert space. 
Radiative transitions, allowed only between subsequent states in the discrete energy ladder, are usually characterised by dipole matrix elements of order $\sqrt{N}$, except for the states close to half-filling, where instead the matrix element becomes of order $N$. 
As initially noticed by Dicke \cite{Dicke54}, the emission thus speeds up, leading to the occurrence of a superfluorescent peak midway through the radiative cascade.

In most physical systems there are also other mechanisms competing with the light-matter interaction. 
While their specific sources and microscopic nature may vary, their effect on the dynamics can generally be described in terms of either nonradiative losses, or energy-conserving pure dephasing.
These mechanisms are detrimental to the development of a coherent, collective dynamics, posing limitations to the observation of cooperative light emission. 
Their study has been the subject of several works that have considered non-ideal conditions for superfluorescence, such as inhomogeneous distribution of the two-level systems resonances, nonradiative emission, radiative emission through a local photonic environment, and effects of the light propagation in large samples that are optically thick \cite{Lehmberg70,Agarwal71,Bonifacio71,Bonifacio71b,Friedberg72,Skribanowitz73,Bonifacio75,Ressayre73,Jodoin74,Ressayre74,Friedberg74,Lee76,MacGillivray76,Schuurmans79,Leonardi82,Leonardi82b,Malcuit87,Maki89,Drummond91,Temnov05,Ishikawa16,Lambert16}. 
Under the continuous pumping condition, several works have also investigated the effects of non-cooperative mechanisms on the cooperative dynamics,
with regard to the properties of light emission \cite{Zheleznyakov89,Belyanin98,Vasilev99,Temnov09,Meiser10a,Meiser10,Bohnet12, Pleinert17,Saez17}, 
and to the related phenomena of optical bistability \cite{Bonifacio78,Gronchi78,bonifacio82,Carmichael86,Sarkar87,Sarkar87b,Amthor15} 
and superradiant phase transitions \cite{Delanty11,Gelhausen16,DallaTorre16,Kirton17,Gegg17a}. 

Here we will exploit a theory able to describe the incoherent interaction of $N$ two-level systems with the electromagnetic field for arbitrary losses and dephasing, in order to gain a microscopic understanding of the trajectory of the system in its evolution away from symmetric states, under the effect of nonradiative losses and pure dephasing. 
Our results shed light on the parameters required to observe superfluorescence in solid-state environments in which the nonradiative loss rate $\gamma_{\text{L}}$ and the pure dephasing rate $\gamma_{\text{D}}$ are usually larger than the spontaneous emission rate of the single two-level system $\gamma_{\text{S}}$. 
In order to accomplish this, we will investigate the different dynamics generated by both phase-breaking mechanisms, and their interplay with radiative emission. 
We begin by introducing a master-equation description of the quantum dynamics. 
After testing the reliability of the theory against exact simulations of the system dynamics for small $N$, we use it to investigate a wide parameter range for large $N$. 
In the regime relevant for solid-state cavity quantum electrodynamics, in which collective strong coupling is observable notwithstanding dephasing effects, $N\gamma_\text{S}\gg\gamma_\text{D}\gg\gamma_\text{S}$, we show that the main effect of dephasing and losses  is that of making superfluorescence sub-optimal, with only a limited amount of the energy released superradiantly. 
Nonetheless, the delay time of pure superfluorescence, $t_\text{d}$, remains a good predictor of the occurrence of the superfluorescent peak. 
Subradiance dominates the later stage of the dynamics, occurring on a timescale set by the phase-breaking mechanisms. 

Finally, we will apply the developed theory to the dilute regime, in which only few excitations are present.
Such a regime, very different from the half-filling relevant to study superfluorescence, is
well described by a linear bosonic theory, extensively used in solid-state physics \cite{Holstein40,Ciuti05,Agranovich09,DeLiberato09b,Kyriienko13}. Specialising our master-equation approach to this regime, we will recover a theory of interacting bright and dark bosonic quasiparticles \cite{DeLiberato09b,Emary13}, in which dephasing and losses are described in terms of intermode scattering rates and particle lifetimes.

\section{General theory}
\label{s0}
In this section we will develop the general theory to study the evolution of a set of $N$ two-level systems evolving under the effect of different incoherent interactions with their environment. Following Dicke \cite{Dicke54} we begin by describing the $n$th two-level system with the algebra of a spin $\frac{1}{2}$. We thus define the operators over its two-dimensional Hilbert space $J_{\xi,n}$, $\xi\in\{x,y,z,+,-\}$, obeying the angular momentum algebra 
\begin{eqnarray}
\lbrack J_{x,n},J_{y,n'} \rbrack&=&iJ_{z,n}\delta_{n,n'},
\end{eqnarray}
and cyclic permutations, with $J_{\pm,n}=J_{x,n}\pm iJ_{y,n}$.
We also define the corresponding total spin operators
$J_{\xi}=\sum_{n=1}^N J_{\xi,n}$.
The free Hamiltonian describing $N$ identical and independent two-level systems of energy $\hbar\omega_0$ can then be written in terms of spins as
\begin{eqnarray}
H_0&=&\hbar\omega_{0}\sum_{n=1}^N J_{z,n}=\hbar\omega_{0} J_{z}
\label{h00}
\end{eqnarray}
and their interaction Hamiltonian with the electric field $\mathbf{E}$ in the dipolar approximation
\begin{eqnarray}
H_{\text{int}}&=&\mathbf{d}\cdot\mathbf{E} \sum_{n=1}^N (J_{+,n}+J_{-,n})=\mathbf{d}\cdot\mathbf{E}  (J_++J_-),
\label{hint}
\end{eqnarray}
where $\mathbf{d}$ is the dipolar moment of each spin.
From the theory of addition of angular momenta we can diagonalise the free Hamiltonian $H_0$ on the basis indexed by the quantum numbers $j$, $m$, and $\alpha$, where 
\begin{equation}
{J}^{2}\ket{j,m,\alpha}=j(j+1)\ket{j,m,\alpha},\quad \quad {J}_{z}\ket{j,m,\alpha}=m\ket{j,m,\alpha},\quad \quad
J_{\pm}\ket{j,m,\alpha}=\sqrt{(j\mp m)(j\pm m+1)}\ket{j,m\pm1,\alpha},
\label{lma}
\end{equation}
with $j=\frac{N}{2}, \frac{N}{2}-1, ..., j_\text{min}+1, j_\text{min}$, and $j_\text{min}=0,\frac{1}{2}$ for $N$ even or odd, respectively, and $m$ an integer or semi-integer, respectively, such that $|m|\leq j$. 
The index $\alpha$, referred to as a symmetry parameter, runs over subspaces with the same value $j$
\cite{Tavis68}.
As clear from \Eq{lma}, the collective spin operators cannot change $\alpha$ and we will thus not mark it explicitly in the following. 
This choice introduces a degeneracy for the Dicke state $\ket{j,m}$ that can be easily computed.
Notice that there are $N$ general states with one excitation, but the ground state $\ket{\frac{N}{2},-\frac{N}{2}}$ is connected only to a single one-excitation state, the maximally-symmetric state $\ket{\frac{N}{2},-\frac{N}{2}+1}\propto J_+\ket{\frac{N}{2},-\frac{N}{2}}$. 
This implies that the degeneracy of the state $\ket{\frac{N}{2}-1,-\frac{N}{2}+1}$ is $(N-1)$. 
From \Eq{lma} we see that each state $\ket{\frac{N}{2}-1,m}$ has the same degeneracy, and is said to belong to the same Dicke ladder.
Analogously, the degeneracy of all of the other inequivalent subspaces can be iteratively computed, leading to the formula for the degeneracy of each Dicke state $\ket{j,m}$ to be
\begin{eqnarray} 
D_{j}&=&\frac{N!(2j+1)}{(\frac{N}{2}+j+1)!(\frac{N}{2}-j)!}.
\label{Dj}
\end{eqnarray}
The degeneracy for a given eigenvalue $\hbar\omega_0m$ of \Eq{h00} instead is \cite{Mandel}
\begin{eqnarray} 
d_m&=&\left(\begin{tabular}{c}$N$\\ $\frac{N}{2}+m$\end{tabular}\right)=\frac{N!}{\left(\frac{N}{2}+m\right)!\left(\frac{N}{2}-m\right)!}.   
\label{dm}
\end{eqnarray}
The coupling with the electromagnetic field, and with the other reservoirs responsible for phase randomisation and energy loss, can be described  
by introducing the master equation for the density matrix $\rho$ of the system
\begin{eqnarray}
\frac{d}{dt}{\rho}&=&i\omega_{0}[{J_{z},\rho}]+\frac{\gamma_\text{S}}{2}\mathcal{L}_{J_{-}}[\rho]+\frac{\gamma_\text{L}}{2}\sum_{n=1}^{N}\mathcal{L}_{J_{-,n}}[\rho]+\frac{\gamma_\text{D}}{2}\sum_{n=1}^{N}\mathcal{L}_{J_{z,n}}[\rho],
\label{me}
\end{eqnarray}
where the Lindblad operator $\mathcal{L}_{O}[\rho]=2O\rho O^{\dagger}- O^{\dagger}O\rho -\rho  O^{\dagger}O,$
describes the different scattering rates quantified by the spontaneous emission rate $\gamma_\text{S}$, the nonradiative relaxation rate $\gamma_{\text{L}}$, and the pure dephasing rate $\gamma_\text{D}$.
The crucial element to notice in \Eq{me} is that the electromagnetic field couples symmetrically with all the spins, and can thus be expressed in terms of the collective spin operators. This expression is not possible for the other two processes, that instead couple with each individual spin independently. In this sense dephasing and losses are both phase-breaking mechanisms, that randomise the phases between different spins. 
This implies that the spontaneous emission  conserves $j$, leading to an evolution simply described by a radiative cascade through the $(2j+1)$ levels of one of the subspaces indexed by fixed values of $j$ and $\alpha$.  This is a huge simplification, restricting the size of the relevant Hilbert space from $2^N$ to $(2j+1)$, allowing one to solve the problem through direct numerical diagonalisation. 
The other two terms instead, by randomising phases, can couple states with different $j$ and $\alpha$.
The problem in this form cannot be solved exactly and brute-force numerical simulations of the dynamics cannot explore the large-$N$ limit, with a tight bound given by the growth of the Hilbert space as $2^N$. 
We will nevertheless exploit the randomisation of the phase-breaking processes by employing a numerical method that exploits the properties of permutational-invariant density matrices, which scale as ${O}(N^3)$ in general and as ${O}(N^2)$ for special initial states, including Dicke states \cite{Chase08,Baragiola10,Damanet16b}. 
Similarly, the fact that the Lindblad operators are invariant under SU(4) symmetry can be employed to reduce the numerical resources required to solve the dynamics \cite{Hartmann16,Xu13}.   
Hereafter we will initially begin to gain a qualitative understanding of the effect of each one of the terms of \Eq{me}, before developing solid approximations that will allow us to simulate the system dynamics.

\section{Dynamics analysis in the Dicke triangle}
We begin by writing the evolution equations for the expectation values of $J_z$ and $J^2$. Using the shortcuts $\tfrac{d}{dt}\langle {J_z}\rangle=\text{Tr}\left[J_z\,\tfrac{d}{dt}{\rho}\,\right]$, 
$\tfrac{d}{dt}\langle{J^2}\rangle=\text{Tr}\left[J^2\,\tfrac{d}{dt}{\rho}\,\right]$, and $J^2=J^2_z-J_z+J_+J_-$, after some algebra we obtain
\begin{eqnarray}
\label{eq7}
\frac{d}{dt}\langle J_{z} \rangle&=&-\gamma_\text{S}\left(
\langle J^2\rangle-\langle J_z^2\rangle+\langle J_z\rangle
\right)-\gamma_\text{L}\left(\langle J_{z}\rangle+\frac{N}{2}\right),
\\
\frac{d}{dt}\langle {{J}^{2}} \rangle&=&-\gamma_\text{D}\left(\langle {J}^{2} \rangle-\langle J_{z}^{2} \rangle-\frac{N}{2}\right)-\gamma_\text{L}\lbrack \langle{J}^{2}\rangle+(N-1)\langle J_{z}\rangle
+\langle J_{z}^{2}\rangle-N\rbrack.
\nonumber
\end{eqnarray}

\begin{figure}[t!]
\begin{center}
\includegraphics[width=10cm]{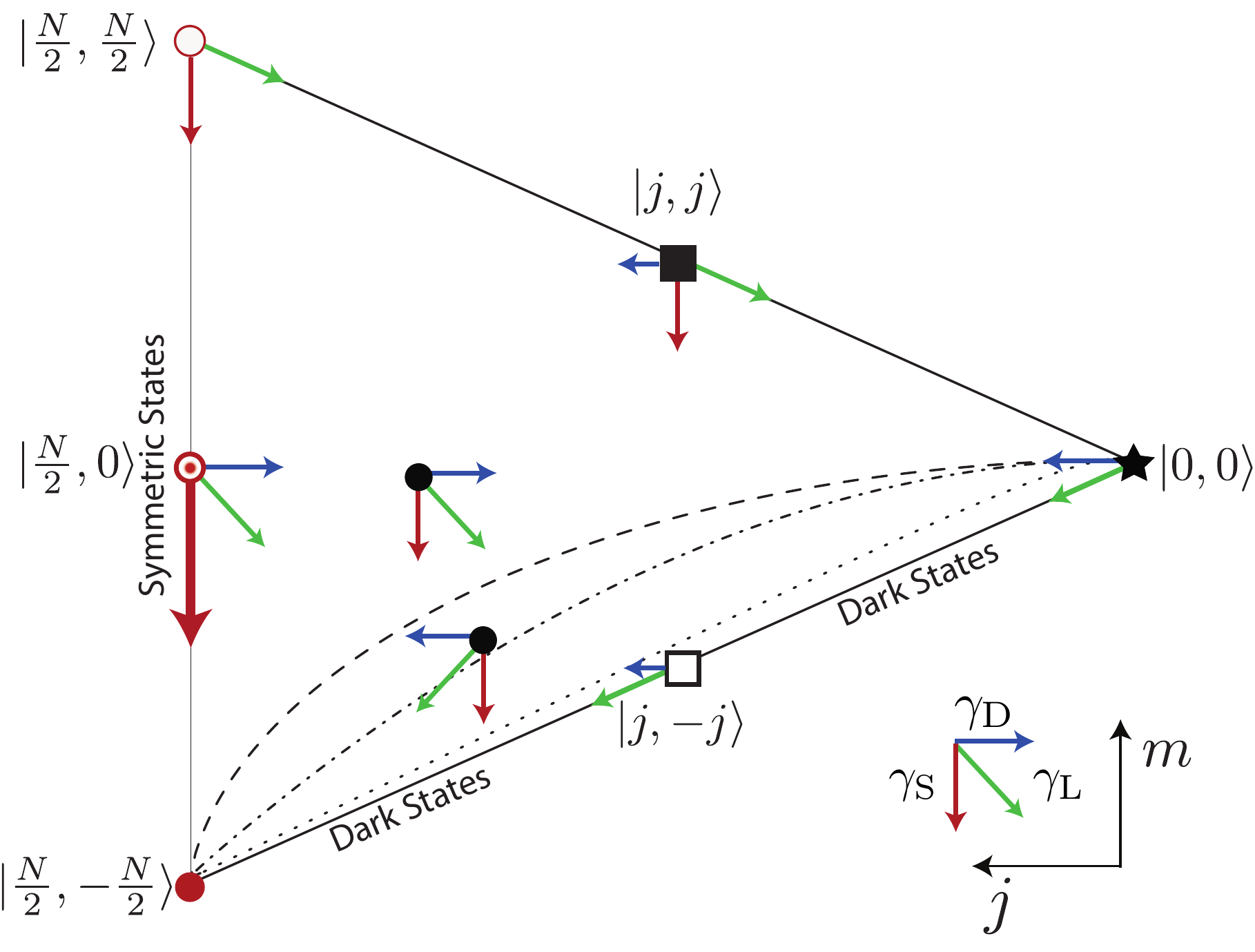}
\caption{\label{fig1}
Hilbert space for $N$ two-level systems parametrised by their energy ($m$, vertical axis) and cooperativity ($j$, horizontal axis). 
The left vertical side of the triangle marks the Dicke ladder of symmetric Dicke states $\ket{\tfrac{N}{2},m}$.
The diagonal sides delimit the borders for the Dicke states $\ket{j,\pm j}$.  
The arrows represent the effect of radiative decay (red, vertical), pure dephasing (blue, horizontal), and nonradiative decay (green, diagonal), on different initial states and their size shows their relative strengths. 
The black segments of the outer triangle mark the border between areas with positive (right) and negative (left) derivative of $j$ for $\gamma_\text{L}/\gamma_\text{D}=10$ (dashed), $\gamma_\text{L}/\gamma_\text{D}=1$ (dot-dashed), and $\gamma_\text{L}/\gamma_\text{D}=0.1$ (dotted).}
\end{center}
\end{figure}

We are now able to analyse the effect of the three scattering channels on an arbitrary state $\ket{j,m}$. 
Projecting \Eq{eq7} on such a pure state we obtain
\begin{eqnarray}
\frac{d}{dt}m&=&-\gamma_\text{S}\left(j^2+j-m^2+m\right)-\gamma_\text{L}\left(m+\frac{N}{2}\right),
 \label{estx}\\
\frac{d}{dt}j&=&-\gamma_\text{D}\frac{j^2+j - m^{2}-\frac N 2}{2j+1}
 -\gamma_\text{L}\frac{j^2+j+(N-1)m+m^{2}-N}{2j+1}.\nonumber
\end{eqnarray}
We will represent the Hilbert space as done in Fig.~\ref{fig1}, in the form of an isosceles triangle, with $m$ on the y-axis and $j$ on the x-axis. 
Spontaneous emission, conserving $j$ and reducing $m$ by one, can only couple a state with the one immediately below it, and indeed in \Eq{estx} only $\tfrac{d}{dt}{m}$ depends on $\gamma_\text{S}$. 
In particular we obtain that the photonic emission rate from a state $\ket{j,m}$ is
\begin{eqnarray}
\frac{d}{dt}{m}_\text{S}&=&-\left(j^{2}+j-m^{2}+m\right)\gamma_\text{S}.
\label{gsr}
\end{eqnarray}
The presence of many spins translates into a state-dependent superradiant enhancement of the spontaneous emission rate. 
From \Eq{gsr} we see that when $m$ is of order $j$, the squares tend to cancel, leading to a contribution at most of order $j$. Only when $m^2 \ll j^2$ the $j^2$ term tends instead to dominate, leading to a superfluorescent emission rate of order $j^2$. In the literature of Dicke superradiance, $j$ is usually named cooperation number or cooperativity: it indeed measures the effective number of spins that coherently participate in the superradiant emission process, leading symmetric states to a burst of peak intensity proportional to $N^2$. 
The effect of spontaneous emission is illustrated in Fig.~\ref{fig1} by red vertical arrows with sizes corresponding to the emission rate.  
This result validates{ the} intuitive picture of superfluorescence: after the system is prepared in the totally-excited state in the top left corner of the triangle in Fig.~\ref{fig1}, it starts emitting photons, moving down and accelerating. 
When approaching $m=0$ it quickly emits most of its energy in a superfluorescent burst, then slowing down until it reaches the ground state in the bottom left corner. 

The pure dephasing term in \Eq{me} on the contrary, cannot change the value of $m$ because its jump operators $J_{z,n}$ commute with $J_z$. It can only modify the value of the cooperativity $j$ as
\begin{eqnarray}
\label{jgD}
\frac{d}{dt}{j}_\text{D}&=&-\frac{j^2+j-m^2-\frac{N}{2}}{2j+1}\gamma_\text{D}.
\end{eqnarray}
Its elastic, phase-randomising effect is shown as blue horizontal arrows in Fig.~\ref{fig1}. Contrary to spontaneous emission, the effect of pure dephasing cannot lead to a superfluorescent enhancement of order $N^2$ due to the denominator in \Eq{jgD}, which limits  
$\tfrac{d}{dt}{j}_\text{D}$ to be at most of order $N$. 
Finally, the nonradiative losses play a role on both $m$ and $j$ \cite{Lee76}
\begin{eqnarray}
\label{jmnr} 
\frac{d}{dt}{m}_\text{L}&=&-\left(m+\frac{N}{2}\right)\gamma_\text{L},\\
\frac{d}{dt}{j}_\text{L}&=&-\frac{j^2+j+(N-1)m+m^2-N}{2j+1}\gamma_\text{L},\nonumber
\end{eqnarray}
and they are represented by green diagonal arrows in Fig.~\ref{fig1}. 
Both contributions are at most of order $N$, implying again the absence of a collective enhancement to the loss rate. 
The previous results are summarised in Table~\ref{t1} for some relevant points of the Hilbert space.
The table prompts a comment on the feasibility of state preparation beyond the case of initial full excitation. Preparing a large ensemble into the maximally-entangled superradiant state $\ket{\frac{N}{2},0}$ requires local control on the two-level systems, yet a
$\tfrac{\pi}{2}-$pulse on the ground state initialises the system in a separable state whose dynamical evolution can be similarly superradiant \cite{Bonifacio71b,Shammah}. 
Regarding the subradiant states $\ket{j,-j}$, theoretical investigations have recently focused on the effect of a photonic cavity \cite{Gegg17a} and on the one-excitation subspace, setting thus $j=\tfrac{N}{2}-1$ and generalizing the formalism of Dicke states to the large sample regime \cite{Scully06,Scully15b,Vetter16}. This is the one that can be experimentally investigated in large atomic clouds \cite{Piovella12}. 
 Yet enhanced state-preparation protocols and locally-controlled dynamics can be obtained in artificial atoms in solid state, especially in circuit QED systems, which have demonstrated to be a promising platform to scale up the ensemble size \cite{vanLoo13,Mlynek14,Kakuyanagi16}.

From \Eq{eq7} we can easily ascertain that $\tfrac{d}{dt}{m}\leq0$, as expected, due to the dissipative nature of the system. On the contrary, $\tfrac{d}{dt}{j}$ can be positive or negative depending on the value of both $j$ and $m$, and the ratio $\gamma_\text{L}/\gamma_\text{D}$. 
Indeed, the possibility of having a dephasing-driven increase of $j$ is not surprising in a finite-dimensional Hilbert space, where quantum revivals can be observed \cite{Haroche}. 
In particular we can understand it by noticing that the ground state corresponds to $j=\frac{N}{2}$ and thus dissipative phenomena that initially tend to reduce the value of the cooperativity have eventually to increase it. 
In Fig.~\ref{fig1} we mark the boundary that separates the region of $\tfrac{d}{dt}{j}>0$ (lower-right concave area) and that of $\tfrac{d}{dt}{j}<0$ (upper-left convex area), showing that for increasing values of $\gamma_\text{L}/\gamma_\text{D}$, the region with positive derivative increases. 
This highlights the fact that the interplay of the two mechanisms can effectively steer the dynamics and could be used to access different points of the Dicke triangle.

From the previous arguments we can qualitatively predict how the system will behave upon initial excitation in the totally-excited state $\ket{\frac{N}{2},\frac{N}{2}}$. If the radiative decay is the faster process, $\gamma_\text{S} \gg \gamma_{\text{L}},\gamma_{\text{D}}$, the system will move mainly downward, resulting in the emission of a superfluorescent burst, although at a reduced intensity with respect to pure fluorescence, as the phase-breaking processes reduce the cooperation number $j$ from its maximum value of $\frac{N}{2}$.  
If the opposite condition $N\gamma_\text{S} \ll \gamma_{\text{L}},\gamma_{\text{D}}$ is valid, then most excitations will be lost through nonradiative decay and the system will thus arrive to the ground state $\ket{\frac{N}{2},-\frac{N}{2}}$ hopping down states lying on, or close to, the two diagonal sides of the triangle in Fig.~\ref{fig1}.
In the intermediate regime, $\gamma_\text{S} \ll \gamma_{\text{L}},\gamma_{\text{D}}$ but $N\gamma_\text{S} \gg \gamma_{\text{L}},\gamma_{\text{D}}$, the cooperation number will initially decrease as excitations are nonradiatively lost, but as the system approaches the middle of the Dicke triangle, cooperative light emission eventually kicks in leading to a superfluorescent burst.

%
\begin{table}[t!]
\begin{center}
 \begin{tabular}{|c|c|c|c|c|c|c}
\hline
&&&&&
\\
& $\bigcirc$& $\blacksquare$&$\odot$ & $\bigstar$& $\square$\\
&&&&&\\
& $\ket{\tfrac{N}{2},\tfrac{N}{2}}$&$\ket{\tfrac{N}{4},\tfrac{N}{4}}$&  $\ket{\tfrac{N}{2},0}$& $\ket{0,0}$& $\ket{\tfrac{N}{4},-\tfrac{N}{4}}$\\
&&&&&\\
\hline
&&&&&\\
$\frac{d}{dt}{m}\ \updownarrow$& $-\left(\gamma_\text{S}+\gamma_\text{L}\right)N$&$-\left(\frac{\gamma_\text{S}}{2}+\frac{3}{4}\gamma_\text{L}\right)N$&$- \frac{\gamma_\text{S}}{4}N^{2}-\frac{\gamma_\text{L}}{2}N$&$-\frac{\gamma_\text{L}}{2}N$&$-\frac{\gamma_\text{L}}{4}N$\\
&&&&&\\
$ \frac{d}{dt}{j} \ \leftrightarrow $& $-\gamma_\text{L}N$ & $\frac{\gamma_\text{D}}{2}-\frac{3}{4}\gamma_\text{L}N$&$-\left(\frac{\gamma_\text{D}}{4}+\frac{\gamma_\text{L}}{4}\right)N$&$\left(\frac{\gamma_\text{D}}{2} +\gamma_\text{L}\right)N$&$\frac{\gamma_\text{D}}{2}+\frac{\gamma_\text{L}}{4}N$\\
&&&&&\\
\hline
 \end{tabular}
\caption{\label{t1}The table shows the derivatives of $\tfrac{d}{dt}{m}$ and $\tfrac{d}{dt}{j}$ in some characteristic points of the Dicke space (see Fig.~\ref{fig1} for a reference to the symbols) according to \Eq{estx}. 
In the Dicke triangle, shown in Fig.~\ref{fig1}, $\tfrac{d}{dt}{m}$ affects the vertical component of the system dynamics and $\tfrac{d}{dt}{m}$ the horizontal one. 
For each rate $\gamma_\text{S}$, $\gamma_\text{L}$, and $\gamma_\text{D}$ the change is given to the leading order in $N$. 
Notice that only the spontaneous emission of the superradiant state $\ket{\tfrac{N}{2},0}$ is collectively enhanced $(\propto N^2)$.}
\end{center}
\end{table}


\section{Quantum dynamics through the truncated hierarchy}
The system given by \Eq{eq7} is not closed, as it contains the average of $J_z^2$. While we could write the equation of motion for its expectation value, it will in turn depend on the evolution of products of three operators and so on, leading to a hierarchy of $O(N^{2})$ coupled equations \cite{CarmichaelI,CarmichaelII}. 
One usual way to deal with this kind of problem, ubiquitous in the study of nonlinear quantum systems, 
is to truncate the hierarchy at a certain order, factorising all the products of more than a given number of operators. The system in \Eq{estx}, in which we calculated the scattering rates from a specific $\ket{j,m}$ eigenvector, effectively corresponds to a first-order approximation, $\langle J_z^2\rangle\simeq\langle J_z\rangle\langle J_z\rangle$.

With the final objective to calculate the quantum dynamics of the system under losses and dephasing, here we will check the solidity of such an approximation. In order to do this we will compare the first-order approximation of \Eq{eq7} both to an exact simulation for up to $N=50$, obtained by a permutational invariant method \cite{Chase08} requiring only ${O}(N^2)$ resources to model the dynamics, and to a second-order approximation for large $N$.
To go to the second order, we complete the system in \Eq{eq7} with the equation for the expectation value of $J_z^2$ 
\begin{eqnarray}
\frac{d}{dt}\langle J_{z}^{2} \rangle&=&\gamma_\text{S}\left(\langle {J}^{2} \rangle+\langle J_{z} \rangle-3\langle J_{z}^{2} \rangle+2\langle J_{z}^{3}\rangle-2\langle J_{z}{J}^{2} \rangle \right)
-\gamma_\text{L}\big[\left(N-1\right)\langle J_{z}\rangle+2\langle J_{z}^{2}\rangle-\tfrac{N}{2}\big], 
\end{eqnarray}
and then factorise the products of three operators assuming
$\langle J_{z}^{3}\rangle\simeq\langle J_{z}\rangle\langle J_{z}^{2}\rangle$ and $\langle J_{z}{J}^{2} \rangle\simeq\langle J_{z}\rangle\langle{J}^{2} \rangle$, leading to 
\begin{eqnarray}
\label{fact}
\frac{d}{dt}\langle{J_{z}^{2}} \rangle&=&\gamma_\text{S}\left(\langle {J}^{2} \rangle+\langle J_{z} \rangle-3\langle J_{z}^{2} \rangle+2\langle J_{z}\rangle\langle J_{z}^{2}\rangle-2\langle J_{z}\rangle\langle{J}^{2} \rangle \right)
-\gamma_\text{L}\big[\left(N-1\right)\langle J_{z}\rangle+2\langle J_{z}^{2}\rangle-\tfrac{N}{2}\big].
\end{eqnarray}
Thanks to the factorisation now \Eq{eq7} and \Eq{fact} form a closed, nonlinear system that can be numerically solved with the proper initial conditions.
With the aim of allowing an easy extension of the factorisation scheme to higher orders, here we report also the general evolution equation for any product of collective operators $J_{+}^{p}J_{z}^{r}J_{-}^{q}$ \cite{CarmichaelI} with $p,q,r\in \mathbb{N}+\{0\}$, generated by \Eq{me} 
\begin{widetext}
\begin{eqnarray}
\frac{d}{dt}\langle{ J_{+}^{p}J_{z}^{r}J_{-}^{q}}\rangle&=&-i\omega_{0}(q-p)\langle{J_{+}^{p}J_{z}^{r}J_{-}^{q}}\rangle+\gamma_{\text{D}}\lbrack -\tfrac 1 2 (p+q)\langle J_{+}^{p}J_{z}^{r}J_{-}^{q}\rangle+pq\langle J_{+}^{p-1} (J_{z}-1)^{r}(\tfrac N 2+J_{z})J_{-}^{q-1}\rangle\rbrack\nonumber\\
&&+\gamma_\text{S}\{\langle J_{+}^{p+1} J_{z}^{r}J_{-}^{q+1}\rangle-\langle J_{+}^{p+1} (J_{z}+1)^{r}J_{-}^{q+1}\rangle+(p+q)\langle J_{+}^{p} J_{z}^{r+1}J_{-}^{q}\rangle+\tfrac 1 2 \left[p(p-1)+q(q-1)\right]\langle J_{+}^{p}J_{z}^{r}J_{-}^{q}\rangle \}\nonumber\\
&&+\gamma_\text{L}\lbrack\langle J_{+}^{p} (J_{z}-1)^{r}(J_{z}+\tfrac N 2)J_{-}^{q}\rangle-\tfrac 1 2(p+q+N)\langle J_{+}^{p}J_{z}^{r}J_{-}^{q}\rangle -\langle J_{+}^{p} J_{z}^{r+1}J_{-}^{q}\rangle\rbrack.\label{dot0}
\end{eqnarray}
\end{widetext}
\begin{widetext}
\begin{figure}[t!]
\begin{center}
\includegraphics[width=16cm]{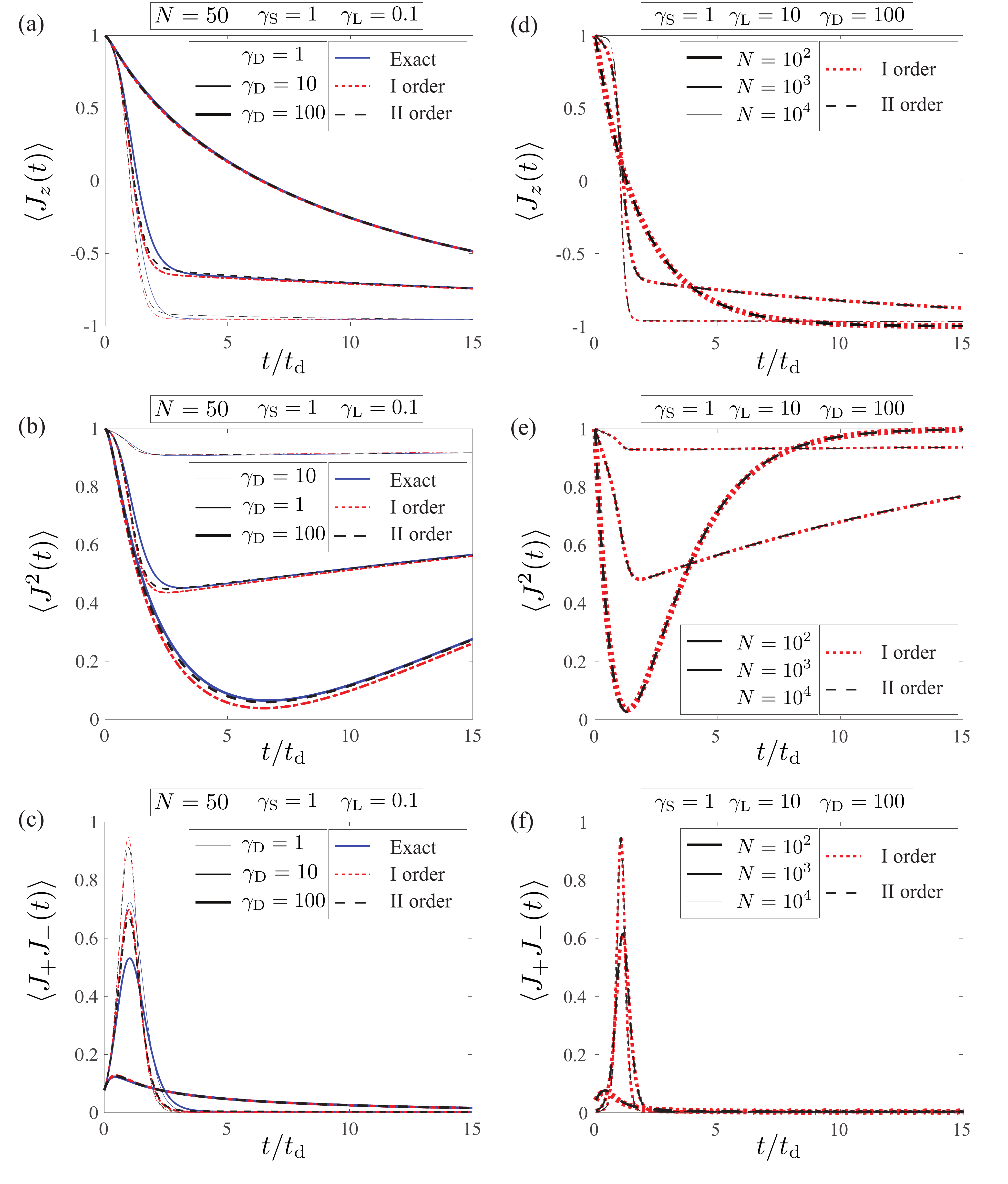}
\caption{\label{fig2}
Left panels: Exact and approximated observables' time evolution for small $N$. The evolution of $\langle{J_{z}(t)}\rangle$ (a) in units of $\frac{N}{2}$, and of $\langle{J}^{2}(t)\rangle$ (b) and $\langle{J}_{+}{J}_{-}(t)\rangle$ (c) in units of $\frac{N}{2}(\frac{N}{2}+1)$ as functions of $t/t_\text{d}$ for $N=50$. We set $\gamma_\text{L}=0.1$, $\gamma_\text{S}=1$ and choose different dephasing values $\gamma_\text{D}=1$ (thin curves), 10 (medium curves), 100 (thick curves). 
The plots show that the nonlinear equations with first-order (red short-dash curves) and second-order (black long-dash curves) approximations agree well
with the exact time evolution (blue solid curves).
Right panels: Approximated observables' time evolutions for large $N$. We plot the first-order (red short-dash curves) with the second-order (black long-dash curves) approximations of $\langle{J_{z}(t)}\rangle$ (d), $\langle{J}^{2}(t)\rangle$ (e), and $\langle{J}_{+}{J}_{-}(t)\rangle$ (f) as functions of $t/t_\text{d}$. 
We set $\gamma_\text{S}=1$, $\gamma_\text{L}=10$, and $\gamma_\text{D}=100$ and vary $N=10^2$ (thick curves), $N=10^3$ (medium curves), $N=10^4$ (thin curves).
}
\end{center}
\end{figure}
\end{widetext}

\section{Numerical study}
\label{s4}
We begin by assessing the validity of the approximations made in the previous Section for small $N$, as in that limit it is possible to solve the dynamics with an exact numerical diagonalisation  \cite{Johansson12,Johansson13}. 
We observe that the Lindblad dynamics of \Eq{me} is permutational-invariant upon exchange of any given couple of two-level systems. 
The time evolution of a Dicke state (or, more generally, any permutational invariant state) will thus be restricted to the subset of other permutational-invariant density matrices, which can be represented with only ${O}(N^3)$ terms, thanks to the fact that matrix elements corresponding to the $D_j$ degenerate subspaces indexed by $\alpha$ are identical. Moreover the density matrix is a block-diagonal matrix in which the coherences $\bra{j,m}\rho\ket{j',m'}$ are non-zero only for $j=j'$ \cite{Chase08}.
Thanks to these facts, the dynamics of \Eq{me} decouples the evolution of coherences $\bra{j,m}\rho\ket{j,m'}$ to that of the populations $\bra{j,m}\rho\ket{j,m}$, so that a set of rate equations for the evolution of populations of the Dicke states requires only ${O}(N^2)$ terms \cite{Chase08,Baragiola10,Damanet16b}. 
This is a considerable simplification, because a brute-force numerical simulation of the Liouvillian superoperator requires $(2^{2N}\times2^{2N})$-dimensional matrices, which would become numerically challenging already for $N$ of the order of 10. By exploiting permutational invariance we can easily explore ensembles of up to $N=50$ two-level systems.  
We point out that there are several approaches to reduce the computational resources related to the dimension of the Lindbladian operator matrices.
Permutational-invariant methods have been applied to the study of the dynamics of spin-squeezing of $N$ spin-$\frac{1}{2}$ particles \cite{Chase08,Baragiola10} and for the purpose of entanglement and state estimation \cite{Moroder12,Novo13}. Permutational invariance has also been used to treat the more general case of multi-level systems dynamics \cite{Gegg16,Gegg17b}.
The dimension of the Liouvillian space can be reduced to ${O}(N^3)$ by exploiting the SU(4) symmetry of the Lindblad superoperators \cite{Hartmann16,Xu13}; indeed in Ref.~\cite{Xu15} a master equation that includes the terms of \Eq{me} has been treated for the study of Ramsey spectroscopy of atomic ensembles.  
Recently these methods have been further developed and they have been employed to study the cooperative effects arising from the interaction with light of an ensemble of $N$ two-level systems both in the context of light emission \cite{Damanet16b,Gong16} and that of superradiant phase transition \cite{Kirton17,Gegg17a}. With regard to the study of driven-dissipative steady states in open systems other tools from condensed matter theory have also been applied \cite{Sieberer16,DallaTorre16,Scarlatella16}.

In Fig.~\ref{fig2}(a)-(c) we show, for $N=50$, the time evolution of the normalised values of $\langle{J}_{z}(t)\rangle$, $\langle{J}^{2}(t)\rangle$, and $\langle{J}_{+}{J}_{-}(t)\rangle$ calculated exactly (blue solid curves), with the first-order approximation of \Eq{eq7}, setting $\langle{J}_{z}^2\rangle=\langle{J}_{z}\rangle^2$ (red short-dash curves), and with the second-order approximation of \Eq{eq7} and \Eq{fact} (black long-dash curves). 
We set $\gamma_\text{S}=1$, $\gamma_\text{L}=0.1$, and tune $\gamma_\text{D}=1$ (thin curves), 10 (medium curves), and 100 (thick curves).
One can observe that qualitatively the first- and second-order approximations match well the exact dynamics, with the second-order faring better than the first-order. 
For $\gamma_\text{D}=1$, both dephasing and losses are small compared to the superradiant rate $N\gamma_\text{S}$.
This is the condition closest to the standard case of pure superfluorescence, $\gamma_\text{D}=\gamma_\text{L}=0$, and we report it as a reference point or benchmark. 
In absence of dephasing and losses the peak of the superradiant pulse would occur after the delay time
\begin{eqnarray}
\label{td0}
t_\text{d}&=&\frac{\ln(N)}{N\gamma_\text{S}},
\end{eqnarray} 
which is calculated by linearizing \Eq{gsr} \cite{Gross82}.
As seen in Fig.~\ref{fig2}(a) (thin curves), for $\gamma_\text{D}=1$ the ensemble's energy is lost on the scale of the superradiance delay time $t_\text{d}$ and, as shown in Fig.~\ref{fig2}(b), the cooperativity of the system is almost constant, marking the fact that the atomic correlations enhance the light emission intensity, which is $\propto N^2$, as seen in Fig.~\ref{fig2}(c). 
For $\gamma_\text{D}=10$ (medium curves), still most of the dynamical changes are occurring on the scale of $t_\text{d}$, yet the cooperativity, shown in Fig.~\ref{fig2}(b), is not constant. 
This is an example of sub-optimal superfluorescence, for which radiative and phase-breaking mechanisms compete ($\gamma_\text{D}=\gamma_\text{S}$), yet the collective enhancement is greater than the other processes $N\gamma_\text{S}\gg \gamma_\text{D}$.
For $\gamma_\text{D}=100$ (thick curves) the system's evolution is dominated by the phase-breaking processes and one retrieves that the dynamics is simply set by the characteristic time $t_0=(\gamma_\text{S}+\gamma_\text{L})^{-1}$ of exponential decay, as shown in Fig.~\ref{fig2}(c). 

In Fig.~\ref{fig2}(d)-(f) we compare the results for the time evolutions for large $N$, with $N=10^2$ (thick curves), $N=10^3$ (medium curves), and $N=10^4$ (thin curves). We can access the large-$N$ limit thanks to the first-order (red short-dash curves) and second-order (black long-dash curves) approximations deriving from \Eq{eq7}, which become almost perfectly overlapped.
This is not surprising considering that the factorisation schemes, as mean field approximations, are expected to improve with increasing $N$. 
Since in the solid state, dephasing is generally the fastest mechanism, we have set now radiative decay as the weakest one, $\gamma_\text{S}=1$, $\gamma_\text{L}=10$, and $\gamma_\text{D}=100$.
Notice that for $N=10^2$ (thin curves) we have $\gamma_\text{D}=N\gamma_\text{S}$, the dynamics is completely incoherent and the decay is exponential with characteristic time $t_{0}$. 
As $N$ is increased to $N=10^3$ and $N=10^4$ (medium and thick curves), the dominant dynamic drive is that of collective decay, since $N\gamma_\text{S}\gg \gamma_\text{D}, \gamma_\text{L}$ even if the system is still in the limit of strong dephasing, $\gamma_\text{D}\gg\gamma_\text{L}\gg\gamma_\text{S}$. 

We are now in a position to study numerically the dynamics for large $N$ and investigating how superfluorescence is affected by losses and dephasing.
We can define operatively a superfluorescence effective delay time $t_\text{d}^\text{eff}$ in the presence of phase-breaking mechanisms as the time at which the system reaches half-filling, $\langle J_{z}(t_\text{d}^\text{eff})\rangle=0$.  
In the limiting case in which $\gamma_\text{D}=\gamma_\text{L}=0$, $t_\text{d}^\text{eff}$ corresponds to the pure superfluorescence delay time $t_\text{d}$ of \Eq{td0}, while when the incoherent mechanisms dominate the dynamics, this is simply the characteristic time $t_{0}$ of the exponential decay.

\begin{figure}[h]
\includegraphics[width=17cm]{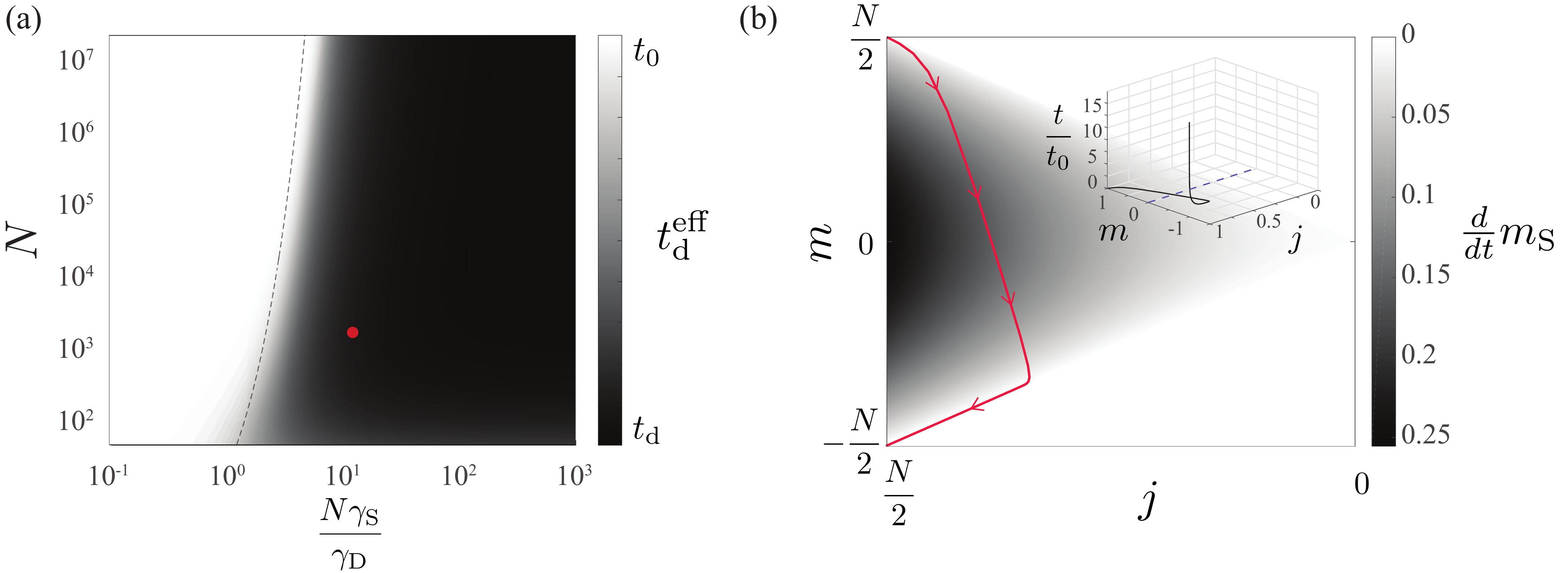}
\caption{\label{fig2coop}
(a) Calculated delay time $t_\text{d}^\text{eff}$ in the parameter space given by $N\gamma_\text{S}/\gamma_\text{D}$ (horizontal axis) and $N$ (vertical axis) for an initially totally-excited ensemble. 
We set $\gamma_\text{S}=1$ and $\gamma_\text{L}=10$. 
The two limiting values $t_0$ and $t_\text{d}$ indicate where the system dynamics is incoherent and where superfluorescent, respectively.  
A dashed curve indicates the approximated analytical expression derived in the literature for the delay in the presence of dephasing. 
The red dot marks a point of the parameter space for which the plot in the right panel shows trajectory and time evolution. 
(b) Trajectory of the system in the $(j,m)$ Dicke space for $N=10^3$, $\gamma_\text{S}=1$, $\gamma_\text{L}=10$, $\gamma_\text{D}=100$. 
The plot shows that superfluorescence can occur even deep inside the Dicke space, far from the $j=\tfrac{N}{2}$ side. 
A color plot shows the value of the photonic emission rate $\tfrac{d}{dt}{m}_{\text{S}}$ of \Eq{gsr} in units of $N^2\gamma_\text{S}$. 
Arrows guide the eye along the trajectory according to the time evolution. 
Inset: The time evolution of the same system is shown explicitly in the Dicke space, with the axes of $m$ and $j$ normalised by $\frac{N}{2}$, and time as the third coordinate. 
The middle of the Dicke triangle, $m=0$ (dashed dark blue line), is reached at a time much shorter than the characteristic time of exponential decay $t_\text{d}^\text{eff}\ll t_0$.}
\end{figure}

In Fig.~\ref{fig2coop}(a) we set $\gamma_\text{S}=1$, $\gamma_\text{L}=10$ and report the value of $t_\text{d}^\text{eff}$ calculated from the closed set of \Eq{eq7} and \Eq{fact} for varying $N$ and $\gamma_\text{D}$.
The threshold between the exponential decay (white background) and the cooperative one (dark shading) is clear, occurring when $N\gamma_\text{S}\gg\gamma_\text{D}$.
This is consistent with previous analyses that estimated the threshold as $T_2^*=\sqrt{\tau_\text{p}t_\text{d}}$, where $T_2^*$ is the critical pure dephasing time, $\tau_\text{p}$ the pulse duration \cite{Leonardi82b,Malcuit87,Maki89,Ishikawa16}, i.e. $\gamma_{\text{D}}^*\simeq \frac{\gamma_\text{S} N}{\sqrt{\ln{N}}}$, plotted as a black dashed curve in Fig.~\ref{fig2coop}(a).
In order to better understand the global dynamics of the system, we can now perform a qualitative analysis of a specific point in the parameter space of Fig.~\ref{fig2coop}(a), choosing one for which superfluorescence occurs but it is relatively near to the threshold. We consider $N=10^3$ and $\gamma_\text{D}=100$ (red dot in panel (a)), and recall that $\gamma_\text{S}=1$ and $\gamma_\text{L}=10$. This means that this point is in the regime $N\gamma_\text{S}>\gamma_\text{D}\gg\gamma_\text{L}>\gamma_\text{S}$, where we expect superfluorescence to occur. 
In Fig.~\ref{fig2coop}(b) we show for this choice of parameters the trajectory of the system in the $(j,m)$ Dicke space (arrows guide the eye to indicate the time evolution). 
The system does not evolve along the left side of the triangular phase space down the longest Dicke ladder for which $j=\tfrac{N}{2}$, as in pure superfluorescence; instead it explores the inner area of the Dicke space until it reaches a sub-optimal dark state $\ket{j,-j}$ on the lower right side of the triangle. 
From then on, as already analysed qualitatively in the previous section, see Table~\ref{t1} for the state $\ket{\tfrac{N}{4},-\tfrac{N}{4}}$, the dynamics is governed by nonradiative energy loss.
The inset of Fig.~\ref{fig2coop}(b) further shows that the two different stages of the dynamics are a first fast one, and a second one, given by the exponential nonradiative decay processes, of order of $t_{0}$. The effect of this two-stage dynamics for the controlled generation of subradiant states has also been analysed in Ref. \cite{Gegg17a} in the regime of small dephasing. 
A color plot of the Dicke space shows the intensity of light emission enhancement in a given point ($j,m$), through $\tfrac{d}{dt}{m}_\text{S}$ from \Eq{gsr} expressed in units of $N^2\gamma_\text{S}$, from which it is clear that although the system does not undergo pure superfluorescence, it crosses the inner superfluorescent area and light emission becomes superradiantly enhanced. 

\section{Bosonisation in the dilute regime}
\begin{figure}[t!]
\begin{center}
\includegraphics[width=10cm]{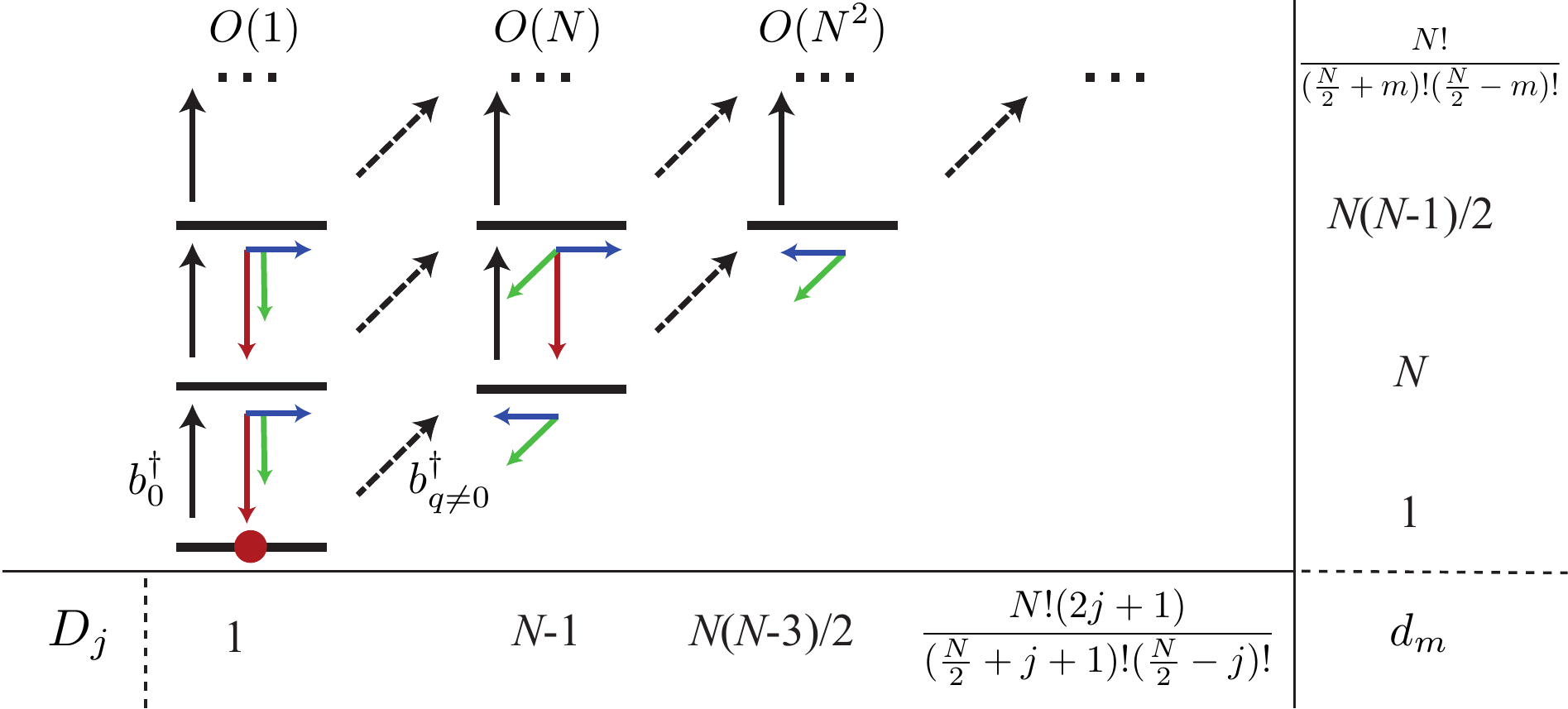}
\caption{\label{figb}
Bosonic approximation valid in the lower part of the full triangular Hilbert space,{ as shown in Fig.~\ref{fig1}}. 
Upward-pointing vertical and oblique arrows represent the bosonic creation operators.  
The bright bosonic operators $b_{0}^\dagger$ connect the Dicke states in the same Dicke ladder, $\Delta m=1$, $\Delta j=0$. 
The dark modes are represented by the dashed arrows $b_{q\neq0}^\dagger$. 
They create an excitation $\Delta m=1$ and in general $\Delta j=0,\pm1$. 
On average, as shown by the rate equation for the dark mode population $\tfrac{d}{dt}{n}_{d}$, in the lower part of the Dicke triangle they destroy cooperativity, $\Delta j=-1$.
The degeneracy of the states in the same Dicke ladder $D_j$ and the degeneracy of the states $d_{m}$ with same energy eigenvalue $m$ are shown. 
As a reference for the dynamics of \Eq{me} and with the bottom panel, also the effect of the scattering rate for the lower excited stated are sketched by a set of small arrows under each state. 
}
\end{center}
\end{figure}

{Having elucidated the impact of losses and dephasing on superfluorescence, here we apply the theory we developed to the lower left corner of the Dicke triangle, characterised by average numbers of excited spins $(m+\frac{N}{2})$ much smaller than their total number $N$.
In this regime the system's excitations can be treated as approximately bosonic.
Such an approach, largely applied in solid-state physics \cite{Ciuti05,Agranovich09,DeLiberato09b,Kyriienko13} is
based on the general idea that a system far from saturation is essentially harmonic. 
It can be formalised through the Holstein-Primakoff transformation that exactly maps the algebra of a spin $j$ into the one of an harmonic oscillator \cite{Holstein40,Ressayre75,Lambert04,Kessler12}
\begin{eqnarray}
\label{holstp}
J_-&=&\sqrt{2j}\sqrt{1-\frac{b^{\dagger}b}{2j}}b,\quad\quad
J_+=\sqrt{2j}b^{\dagger}\sqrt{1-\frac{b^{\dagger }b}{2j}},\quad\quad
J_z=b^{\dagger}b-j,
\end{eqnarray}
with $\left[b,b^{\dagger}\right]=1$.
The number of bosonic excitations in a generic state $\bra{j,m}b^{\dagger}b\ket{j,m}$ is thus equal to
$(j+m)$, that is the number of excitations over the lowest lying state for a fixed value of $j$.
If such a number is much smaller than $j$ (dilute regime) we can then do a lowest-order expansion of the square roots in the small parameter $ b^{\dagger}b/2j$, recovering standard quadratic, bosonic Hamiltonians.
In our case this formal approach is not directly applicable because the mapping explicitly depends on $j$ and phase-breaking effects would thus have an impact on the very definition of the bosonic operators \cite{DallaTorre16}. 
In order to understand how the bosonic approach maps into the full picture of a spin Hamiltonian, we will instead proceed more phenomenologically, studying the structure of the lower left corner of the Hilbert space in Fig.~\ref{figb}, corresponding to the dilute regime.
We begin by formally defining the operators
\begin{eqnarray}
\label{bdef}
b^{\dagger}_p&=&\sum_n f_n^p J_{+,n},\quad p\in\lbrack0,N-1\rbrack,
\end{eqnarray}
where the $f$s form an orthonormal basis  
\begin{eqnarray}
\label{fsym}
\sum_n f_n^p\bar{f}_n^{q}=\delta_{p,q},
\quad\quad
f_n^0=\frac{1}{\sqrt{N}}\quad \forall n.
\end{eqnarray}
Using \Eq{bdef} we can then calculate 
\begin{eqnarray}
\label{jzb}
J_z=\sum_{p}\left(b^\dagger_pb_p-\tfrac{1}{2}\right),
\end{eqnarray}
which stands as a generalisation of the third term of \Eq{holstp}.}
The operators of \Eq{bdef} are defined such that the $N$ states $b^{\dagger}_p\ket{\frac{N}{2},-\frac{N}{2}}$ span the one excitation subspace. 
The bosonic approximation consists in assuming that those modes remain orthogonal and bosonic also in the excited manifolds, that is
\begin{eqnarray}
\label{boscom}
\lbrack b_p,b^{\dagger}_q\rbrack&=&\delta_{p,q}.
\end{eqnarray}
Using \Eq{fsym}, we can rewrite \Eq{hint} in the form
\begin{eqnarray}
H_{\text{int}}&=&\sqrt{N}\mathbf{d}\cdot\mathbf{E} (b_0+b_0^{\dagger}),
\end{eqnarray}
from which we see that  only the $p=0$ (bright) mode couples with light, with a dipole $\sqrt{N}$ times the bare one \cite{Kakuyanagi16}, while the other $N-1$ modes are dark. 
Notice that the collective $\sqrt{N}$ enhancement to the dipole
has an impact on the system's spectrum, allowing us to explore non-perturbative coupling regimes (strong \cite{Kavokin}, ultra \cite{Ciuti05}, or deep \cite{DeLiberato14}), but it does not affect the emission rate. 
Since the emission rate is proportional to the square of the dipole, we recover a total emission that scales as $N$, as if each dipole was emitting independently. 
This is as expected from the discussion of the previous section, in which we saw that an enhancement of the emission rate can only be obtained for a system close to half-filling $\lvert m \lvert \ll j$, where the bosonic approximation does not hold.

In order to investigate the role of losses and dephasing on the dynamics of bright and dark modes in the bosonic approximation, in Fig.~\ref{figb} we plot a zoom of the lowest left corner of the Hilbert space, in which the bosonic approximation is relevant, explicitly showing the degeneracy of each state, $D_{j}$ \cite{Mandel}. 
In the one excitation manifold, which is exactly spanned by the $N$ vectors $b^{\dagger}_p\ket{\frac{N}{2},-\frac{N}{2}}$, the effect of both spontaneous emission and of nonradiative losses is trivial: as both need to lower the number of excitations and there is a single ground state, they will scatter each state into the ground state $\ket{\frac{N}{2},-\frac{N}{2}}$. 
From \Eq{jgD} we can see that the pure dephasing term will lead, to the dominant order in $N$, to $\tfrac{d}{dt}{j}_\text{D}=-\gamma_\text{D}$ for the bright state and to $\tfrac{d}{dt}{j}_\text{D}=\frac{1}{N}\gamma_\text{D}$ for the dark ones. This can be
easily interpreted recalling that the dephasing randomises the phases between the different spins, thus transforming one mode into the other. 
Given that there is a single bright mode and $(N-1)$ dark ones, any phase change will transform a bright mode into a dark one, decreasing $j$, but the majority of the phase changes will not influence the population of the dark modes. This describes well, for example, intersubband transitions in doped quantum wells, where the line width of the bright mode coupled with the electromagnetic field is determined by its dephasing rate, that transforms it into a dark, uncoupled excitation that eventually relaxes nonradiatively \cite{Capasso,Shammah15}.

This picture can be extended to the higher excited manifolds taking into consideration that the addition of a dark excitation in the dilute regime effectively {translates only into} an upward-right diagonal step, as shown in Fig.~\ref{figb}, reducing the value of $j$ by 1 while increasing $m$, due to the greater weight of the degeneracy $D_j$ for the Dicke ladders of lower cooperation number $j$, for $j=\tfrac{N}{2}$.
This can be verified by counting the available modes: there are ${O}(N)$ dark modes and a single bright one, so that there are ${O}(N^2)$ states with two dark excitations and ${O}(N)$ states with a dark and a bright excitation and analogously for higher excitation numbers. 
We are using orders of magnitude here instead of the exact mode counting because considering $N$ independent bosonic modes leads to over-counting the number of states, with an error of the order of the total number of excitations divided $N$. Once again we see that the bosonic approximation breaks down when the number of excitations becomes comparable with the total number of spins.

In order to describe the effect of phase-breaking mechanisms on the multi-excitation dynamics in the bosonic approximation, we start by introducing the bright and dark excitation populations
\begin{eqnarray}
\label{nbnd}
{n}_b&=&\langle b^\dagger_0b_0\rangle,\\
{n}_d&=&\langle\sum_{p\neq0}b^\dagger_pb_p\rangle\nonumber.
\end{eqnarray}
Using \Eq{bdef} and \Eq{fsym} in \Eq{nbnd} we obtain
\begin{eqnarray}
\label{Jzxx}
n_b&=&\frac{1}{N}\lbrack j(j+1)-m^2+m\rbrack, \\
n_d&=& m-\frac{1}{N}\lbrack j(j+1)-m^2+m\rbrack+\frac{N}{2},\nonumber
\end{eqnarray}
that in the dilute regime, to the dominant order in $N$ leads to 
\begin{eqnarray}
\label{Jzxx2}
n_b&=&j+m, \\
n_d&=& \tfrac{N}{2}-j.\nonumber
\end{eqnarray}
The physical content of \Eq{Jzxx2} can be immediately verified, as the total number of excitations is clearly given by $n_b+n_d=m+\tfrac{N}{2}$, whereas the number of dark excitations increases by 1 when reducing $j$ by 1.

We can now capture the dynamics of the populations of bright and dark states in \Eq{nbnd} either directly from \Eq{Jzxx} using \Eq{eq7} or more formally by a derivation in time of \Eq{nbnd} using \Eq{bdef} and \Eq{jzb}, as both give to the dominant order in $N$
\begin{eqnarray}
\label{rateeq}
\tfrac{d}{dt}{n}_b&=&-(N\gamma_\text{S}+\gamma_\text{D}+\gamma_\text{L}) n_b, \\
\tfrac{d}{dt}{n}_d&=&-\gamma_\text{L} n_d+\gamma_\text{D} n_b. \nonumber
\end{eqnarray}
Such a system is consistent with our initial interpretation: All the scattering processes destroy the bright mode, while the dark modes can decay only through nonradiative losses and they are replenished by dephasing that transforms a bright mode into a dark one.
Note that in the second line of \Eq{rateeq}, the absence of a term proportional to $\gamma_\text{L} n_b$ that one would expect from the previous discussion, as the phase-randomising effect of the nonradiative losses, should transform bright into dark modes. 
This effect is absent because from \Eq{jmnr} we see that in the symmetric-states subspace $j=\frac{N}{2}$, $n_b=-\frac{N}{2}+m\ll N$, $\tfrac{d}{dt}{j}_{\text{L}}$ is of order $\frac{1}{N}$. 
That is, in the dilute excitation regime nonradiative losses do not cause dephasing.
An equivalent result could be obtained also by neglecting the jump term $\tfrac{\gamma_\text{L}}{2}\sum_{n}J_{-,n}\rho J_{+,n}$ in the Lindblad superoperator $\mathcal{L}_{J_{-,n}}[\rho]$ in \Eq{me}, a procedure that is justified in the case of weak pumping, which falls into the dilute regime treated here
\cite{Visser95,Brecha99,Saez17}.
\section{Conclusions}
\label{Conclusions4}
We have performed a study of the interplay of superradiant light emission and local phase-breaking mechanisms. Describing dephasing and nonradiative decay with a master equation formalism, solved through a mix of approximate and exact approaches, we investigated how the interplay of the different scattering channels influences the characteristics, and even the occurrence of the superfluorescent burst. Our results set clear requirements to observe superfluorescence in different condensed-matter systems. 
We also show how our treatment, when applied to the dilute excitation regime, allows one to describe phase-breaking mechanisms in term of scattering of bright and dark modes.
\section{Acknowledgements}
We thank Adam Miranowicz, Anton Frisk Kockum, Carlos S\'anchez Mu\~{n}oz, Fran\c cois Damanet, and Peter Kirton for useful discussions and comments. 
S.D.L. acknowledges support from EPSRC Grant EP/M003183/1. 
S.D.L. is a Royal Society Research Fellow.
F.N. was partially supported by the RIKEN iTHES Project,
MURI Center for Dynamic Magneto-Optics via the AFOSR Award No. FA9550-14-1-0040,
the Japan Society for the Promotion of Science (KAKENHI), 
the IMPACT program of JST,
and CREST grant No. JPMJCR1676.
F. N. and N.L. acknowledge support from the RIKEN-AIST Joint Research Fund.
F. N., N.S., and N.L. acknowledge support from the Sir John Templeton Foundation.
N.S. acknowledges support from EPSRC Grant EP/L020335/1.
\bibliography{references5}

\end{document}